\documentclass[conference]{IEEEtran}
\IEEEoverridecommandlockouts
\usepackage{cite}
\usepackage{amsmath,amssymb,amsfonts}
\usepackage{algorithmic}
\usepackage{graphicx}
\usepackage{textcomp}
\usepackage{xcolor}
\usepackage{hyperref}
\usepackage{url}
\usepackage{tabularx}
\usepackage[table]{xcolor} 
\usepackage{float}
\usepackage{booktabs}
\usepackage{multirow}
\usepackage{cleveref}
\usepackage{mathrsfs} 
\usepackage{graphicx} 
\usepackage{subcaption} 
\usepackage{makecell}
\def\BibTeX{{\rm B\kern-.05em{\sc i\kern-.025em b}\kern-.08em
    T\kern-.1667em\lower.7ex\hbox{E}\kern-.125emX}}
\begin{document}

\title{DeepAssert: An LLM-Aided Verification Framework with Fine-Grained Assertion Generation for Modules with Extracted Module Specifications}

\author{
\IEEEauthorblockN{
Yonghao Wang\textsuperscript{1,2},
Jiaxin Zhou\textsuperscript{3},
Hongqin Lyu\textsuperscript{1,2}, 
Zhiteng Chao\textsuperscript{1},
Tiancheng Wang\textsuperscript{1},
and
Huawei Li\textsuperscript{1,2}}
\IEEEauthorblockA{\textsuperscript{1}State Key Lab of Processors, Institute of Computing Technology, CAS, Beijing, China}
\IEEEauthorblockA{\textsuperscript{2}University of Chinese Academy of Sciences, Beijing, China}
\IEEEauthorblockA{\textsuperscript{3}Beijing Normal University}

\IEEEauthorblockA{wangyonghao22s@ict.ac.cn}
\IEEEauthorblockA{202321130074@mail.bnu.edu.cn}
\IEEEauthorblockA{\{lvhongqin24b, chaozhiteng, wangtiancheng, lihuawei\}@ict.ac.cn}
}

\maketitle

\begin{abstract}
Assertion-Based Verification (ABV) is a crucial method for ensuring that logic designs conform to their architectural specifications. However, existing assertion generation methods primarily rely on information either from the design specification, or register-transfer level (RTL) code. The former methods are typically limited to generating assertions for the top-level design. As the top-level design is composed of different modules without module-level specifications, they are unable to generate deep assertions that target the internal functionality of modules. The latter methods often rely on a golden RTL model, which is difficult to obtain. To address the above limitations, this paper presents a novel large language model (LLM)-aided verification framework named DeepAssert. DeepAssert is capable of analyzing the invocation relationships between modules and extracting independent specifications for each module with its I/O port information. These extracted specifications are subsequently used to guide LLMs to automatically generate fine-grained deep assertions for these modules. Our evaluation demonstrates that DeepAssert significantly outperforms existing methods such as AssertLLM and Spec2Assertion in generating high-quality deep assertions for modules. Furthermore, when integrated with these methods, DeepAssert can enhance the overall quality of the assertions generated. This allows for a more comprehensive and effective verification process.
\end{abstract}


\section{Introduction}
Functional verification is essential in integrated circuit (IC) design, where verification engineers check whether the designers' register-transfer level (RTL) code meets the architectural specifications. Assertion-Based Verification (ABV) is widely adopted in RTL design due to its ability to enhance observability and reduce simulation debugging time by up to 50\% \cite{1, 2}. In particular, high-quality SystemVerilog assertions (SVA) for formal property verification (FPV) are critical within ABV methodologies \cite{3,4}, as they should accurately reflect both high-level design intent and low-level RTL details. However, specifications often contain ambiguities \cite{4,6,9}, and different designers may implement the same specification in significantly different ways, which makes translating specifications into effective SVA assertions a time-consuming and labor-intensive process \cite{7}. As design complexity increases, developing more efficient methods for SVA generation has become an urgent necessity.

\begin{figure}
  \centering
  \includegraphics[width=1\linewidth]{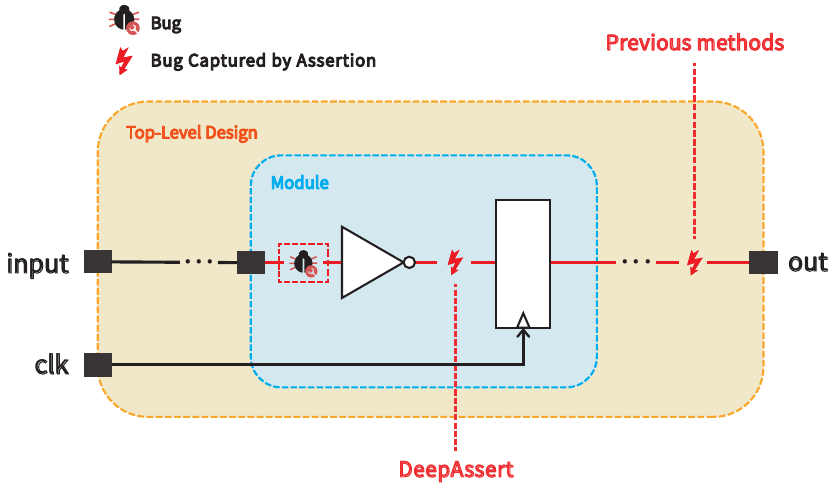}
  \caption{Motivation of DeepAssert: Previous methods generate assertions only for the top-level design, leading to long bug capture latency and difficulty in locating bugs within modules. DeepAssert addresses this by creating fine-grained deep assertions for internal module signals, speeding up bug capture and simplifying bug location.}
  \label{fig:1}
\end{figure}

The rapid development of large language models (LLMs) has inspired new approaches for generating hardware design assertions. Research on using LLMs to automatically generate SVAs can be divided into two main categories. The first category relies solely on LLMs to translate natural language specifications into assertions without using RTL code \cite{10,11,20,23,24}. However, these methods are problematic because they depend on high-level, ambiguous, and incomplete specifications \cite{4,9}, which ignore critical RTL implementation details. The second category integrates both specifications and RTL code \cite{6,12,13,21}, thereby capturing more detailed functional characteristics. However, the assertions generated from the RTL code under verification, may inherit possible logic errors that contradict the specification, thus lose the ability to find errors inside the RTL code. As a result, these methods often rely on a golden RTL model, which is difficult to obtain. To address these limitations, some studies have improved assertion generation accuracy by establishing mappings between specifications and assertions \cite{14} or by fine-tuning mappings between Verilog code and assertions \cite{15}.

However, the existing assertion generation methods are primarily concentrating on extracting high-level functional logic from top-level specifications, neglecting the demand for deep assertions, which refer to fine-grained assertions specifically targeting internal signals within modules. This limitation arises because designers usually decompose the original specification into implicit module specifications mentally, instead of writing it into a module spec file. As a result, the previous methods can only generate coarse-grained SVAs at the top level, failing to provide the fine-grained verification needed for modules. As shown in Fig.\ref{fig:1}, deep assertions are typically closer to the bugs occurring within the modules, which facilitate the debugging process for verification engineers. Therefore, deep assertions are crucial for localizing issues, verifying internal logic, and maintaining module/sub-module functionality.

To address this issue, a major challenge is how to extract the specifications for each module based on the existing original specification and the RTL code. Moreover, given the potential for errors in the RTL code, the extraction process must be independent of the module's RTL implementation details. Interestingly, in real-world verification, the configurations of input and output ports of modules are assumed to be correct; otherwise, verification would be unnecessary. Thus, by analyzing these port configurations and the inter-module invocation relationships within the top-level design, it is possible to infer the module's overall functionality and signal logic descriptions. Given the semantic and contextual understanding capabilities of LLM, this inference is feasible \cite{16}.

In this paper, we propose DeepAssert, a verification framework designed to guide the LLMs to analyse the original high-level specification, module port information, and inter-module invocation relationships. This enables the accurate extraction of module-level specifications without relying on RTL implementation details, thereby facilitating the generation of fine-grained deep assertions for modules.

The contributions of this paper are summarized as follows:

\begin{enumerate}[]
    \item DeepAssert pioneers the approach to accurately extract module-level specifications without relying on implementation details of RTL codes.
    \item To the best of our knowledge, DeepAssert is the first automatic assertion generation method capable of generating finer-grained deep assertions for modules, which can effectively improve the quality of assertion generation and thus reduce the debugging time.
    \item DeepAssert exhibits strong compatibility, and can be seamlessly integrated with any existing automatic assertion generation method.
\end{enumerate}

\section{Preliminaries and Problem Formulation}

\subsection{Assertion Generation Based on LLM}

The development of automated hardware assertion generation frameworks has advanced significantly, leveraging the robust semantic understanding of LLMs. Recent studies have focused on refining these frameworks to improve efficiency and accuracy. Rahul Kande et al. \cite{12} were early explorers of using LLMs for hardware security assertions. Assertllm by Fang et al. \cite{10} was the first method to process complete specification files and generate comprehensive SVAs for each architectural signal. Bai et al. \cite{6} introduced AssertionForge, which constructed a unified Knowledge Graph (KG) from both natural language specifications and RTL code, bridging the gap between high-level intent and low-level details to enhance assertion quality. Wu et al. \cite{14} proposed Spec2Assertion, which used progressive regularization and Chain-of-Thought prompting to generate high-quality assertions directly from specifications, achieving superior syntactic correctness and functional relevance.

\subsection{Problem Formulation}

In the first step, We denote the original specification files as $\mathcal{S}$ and the corresponding golden RTL code as $\mathcal{R}$. Our module relationship extraction process, denoted as ${RE}$, is designed to extract relevant inter-module relationships from $\mathcal{R}$. This relationship information includes the inter-module invocation relationship ($\mathcal{L}^r$), module port information ($\mathcal{L}^p$), and signal propagation changes ($\mathcal{L}^s$). The relationship extraction process can be formally expressed as follows: \\
\textbf{Problem 1:} (Inter-Module Relationships Extraction)
\begin{equation}
RE\left(\mathcal{R}\right) \rightarrow \{\mathcal{L}^r, \mathcal{L}^p, \mathcal{L}^s\}
\end{equation}

In the second step, for ease of reference, we define $\mathcal{L} := (\mathcal{L}^r, \mathcal{L}^p, \mathcal{L}^s)$ to encapsulate the inter-module relationship information. Building upon this, we aim to extract the dedicated specification for each module \( m_{i} \) without relying on the internal implementation details of $\mathcal{R}$. This task is handled by a second stage, denoted $SE$, which leverages both the original high-level specification $\mathcal{S}$ and the inter-module relationship information contained in $\mathcal{L}$. The output of this process is a set of module-level specifications. The process can be formally expressed as follows:\\
\textbf{Problem 2:} (Sub-module Specification Extraction)
\begin{equation}
\forall m_i \in R, SE(\mathcal{S}, \mathcal{L}, m_i) \rightarrow\mathcal{F\text{\((m_i\))}}
\end{equation}
where $\mathcal{F\text{\((m_i\))}}$ represents the inferred functional specification of module \( m_i \).

\begin{figure*}[h]
\centering
\includegraphics[width=1.01\linewidth]{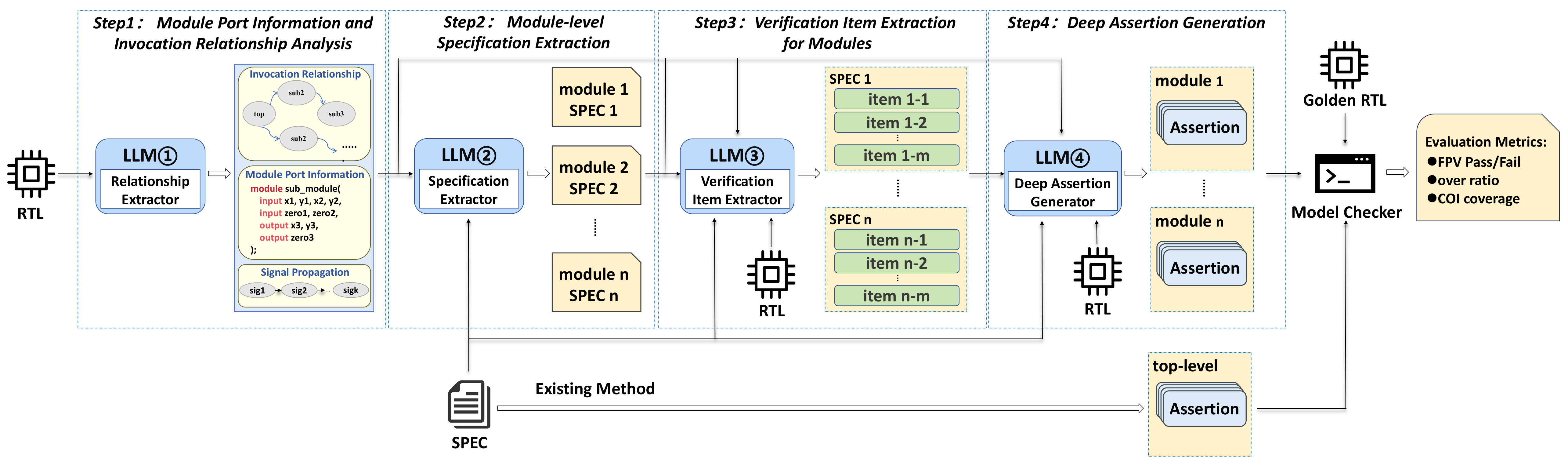}
\caption{The flowcharts of the DeepAssert framework for generating deep assertions and integrating existing methods to generate top-level assertions, and the generated assertions are further evaluated against the golden RTL implementation using model checking tools.}
\label{fig:2}
\end{figure*}

The third step, denoted as $IE$, focuses on identifying key verification items for each module. Each item consists of a condition and its expected result, serving as the basis for generating accurate assertions. The process can be formally expressed as follows: \\
\textbf{Problem 3:} (Verification Item Extraction)
\begin{equation}
\forall m_i \in R, IE\left(\mathcal{R}, \mathcal{S}, \mathcal{L}, \mathcal{F\text{\((m_i\))}} \right) \rightarrow\mathcal{V(\text{\(m_i\))}}
\end{equation}
where $\mathcal{V(\text{\(m_i\))}}$ denotes the set of verification items for module \( m_i \).

We denote our final step as $DGen$ to generate deep assertions for each module. The process can be formally expressed as follows:\\
\textbf{Problem 4:} (Deep Assertion Generation)
\begin{equation}
\forall m_i \in R, DGen(\mathcal{S},\mathcal{R}, \mathcal{L} , \mathcal{F\text{\((m_i\))}}, \mathcal{V(\text{\(m_i\))}}) \rightarrow \mathcal{A(\text{\(m_i\))}}
\end{equation}
where $\mathcal{A(\text{\(m_i\))}}$ denotes the set of generated deep assertions for module \text{\(m_i\)}.

\section{DeepAssert Framework}
    
\subsection{Workflow Overview}\label{AA}

To address the challenge of generating accurate and fine-grained assertions for modules, we propose the DeepAssert framework. DeepAssert extracts dedicated specifications for each module without relying on its RTL code, thereby enabling the generation of targeted and deep assertions for module signals. Specifically, as shown in Fig.\ref{fig:2}, the entire DeepAssert framework is designed as a four-step workflow for automatic deep assertion generation.

\subsection{Step 1: Module Port Information and Invocation Relationship Analysis}

\begin{figure}[h]
  \centering
  \includegraphics[width=0.9\linewidth]{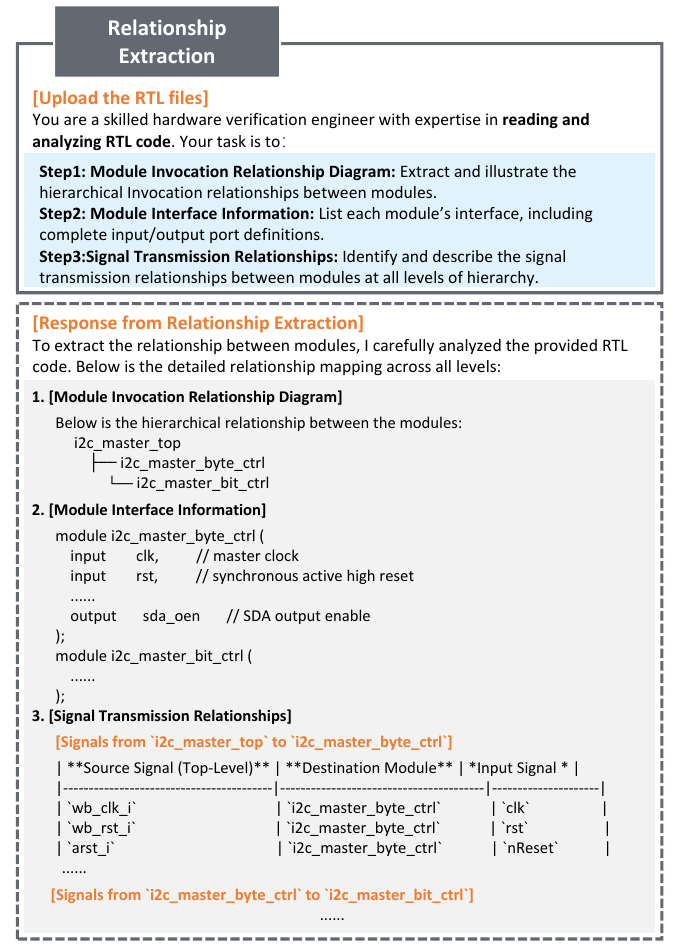}
  \caption{Concise prompt and response of Relationship Extraction.}
  \label{fig:3}
\end{figure}

In practical scenarios, the RTL code provided to verification engineers is assumed to accurately define the input and output ports for each module. Based on this assumption, we can directly extract the inter-module invocation relationship graph, port information, and signal propagation across module interfaces from the RTL code. This extraction allows us to understand the functionality of each module and the overall system behavior. To accurately extract the aforementioned information, our work introduces a dedicated LLM\textcircled{1} \texttt{Relationship Extractor} specifically for extracting the three parts from RTL code. We present the response generated by the LLM\textcircled{1} under guided prompting, as shown in Fig.\ref{fig:3} 

\subsection{Step 2: Module-Level Specification Extraction}

Extracting module-level specifications directly from the RTL code implementation might seem straightforward. However, as discussed in Section I, RTL code may contain errors. Consequently, specifications derived from erroneous RTL implementations are likely to be incorrect as well.

\begin{figure}[h]
  \centering
  \includegraphics[width=0.9\linewidth]{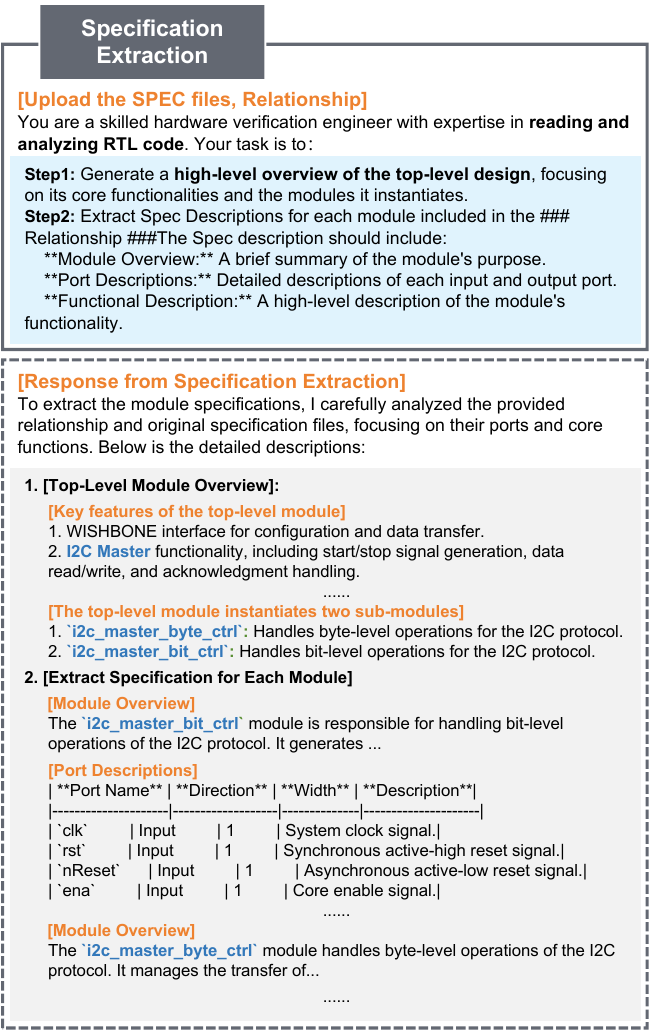}
  \caption{Concise prompt and response of Module-Level Specification Extraction.}
  \label{fig:4}
\end{figure}

To address the aforementioned challenge, we propose an LLM\textcircled{2} \texttt{Specification Extractor} that can accurately extract the specifications for modules without relying on the implementation details of the RTL code. Specifically, the LLM\textcircled{2} first understands the functional overview of the entire design provided in the original specification. Based on this understanding, the LLM\textcircled{2} further leverages the relationships extracted in Step 1 to accurately infer the overall functionality of a module and the behavior of each signal at its ports. Consequently, a detailed module-level specification is generated, as shown in Fig.\ref{fig:4}.

\subsection{Step 3: Verification Item Extraction for Modules}

Based on the module-level specifications extracted in Step 2, we use LLM\textcircled{3} \texttt{Verification Item Extractor} to extract \textbf{verification items—concise, atomic statements of what must be verified directly from the module-level specifications}, an example as shown below: 
\begin{equation*}
\scriptsize
\begin{aligned}
&\texttt{VI-1: 'go' should be high when any of the commands ('start',} \\
&\texttt{'stop', 'read', 'write') are asserted, and 'cmd\_ack' is low.}
\end{aligned}
\end{equation*}
These items guide the LLM to generate a limited number of effective deep assertions in subsequent step.

\subsection{Step 4: Deep Assertion Generation}

Based on the information provided by the previous steps, we propose the LLM\textcircled{4} \texttt{Deep Assertion Generator} that can be guided to accurately generate deep assertions within modules. Furthermore, we have provided a standard template \texttt{\{source\_module\_name\}.\{signal\_name\}} to effectively constrain the generation behavior of LLM\textcircled{4}.

\section{Experiments}

\definecolor{lightblue}{RGB}{70,190,240}
\definecolor{lightgreen}{RGB}{152, 251, 152}

    
\subsection{Experimental Setup}

The benchmark data used in this study are sourced from the open-source dataset in \cite{18}, which comprises 20 designs. Each design includes a specification file and the corresponding golden RTL implementation code. We employ the commercial formal verification tool Cadence JasperGold (version: 21.12.002) for correctness analysis. All experiments are conducted on a server equipped with an Intel(R) Xeon(R) Gold 6148 CPU @ 2.40GHz.

To evaluate the effectiveness of DeepAssert, we conduct comparative experiments with two state-of-the-art (SOTA) assertion generation methods under consistent experimental settings:

\begin{itemize}
    \item \textbf{AssertLLM \cite{10}:} This is a multi-agent method that sequentially extracts signal descriptions from the specification documents, maps them to RTL signals, and uses LLMs to generate assertions, thereby enabling pre-RTL assertion generation.
    \item \textbf{Spec2Assertion \cite{14}:} This method utilizes LLMs equipped with progressive regularization and Chain-of-Thought prompting to formalize and extract functional descriptions, thereby directly generating pre-RTL assertions from design specifications.
\end{itemize}

All  three methods utilize GPT-4o as the experimental model to observe the results.

\subsection{Evaluation Metrics and Benchmarks}

We utilize the metrics provided in JasperGold, while the detailed evaluation criteria is shown in Table \ref{Summary of Evaluation Metrics}. 

\begin{table}[h]
\centering
\renewcommand{\arraystretch}{1.4} 
\caption{Summary of Evaluation Metrics}
\label{Summary of Evaluation Metrics}
\begin{tabular}{c|c}
\hline
\hline
\cellcolor{gray!20}\textbf{Evaluation Metrics} & \cellcolor{gray!20}\textbf{Summary} \\ 
\hline
\makecell{$N$} & \makecell{\hspace*{-25pt} The number of generated SVAs}\\ 
\makecell{$S$} & \makecell{\hspace*{-10pt} The number of syntax-correct SVAs}\\
\makecell{$P$} & \makecell{\hspace*{-18pt} The number of FPV-passed SVAs}\\
\makecell{\textbf{$BFC$}} & \makecell{\hspace*{-14pt} Branch coverage in formal analysis} \\ 
\makecell{\textbf{$SFC$}} & \makecell{\hspace*{-5pt} Statement coverage in formal analysis} \\ 
\makecell{\textbf{$TFC$}} & \makecell{\hspace*{-15pt} Toggle coverage in formal analysis} \\
\hline
\hline
\end{tabular}
\\
\vspace{0.1cm} 
\scriptsize 
\raggedright 
*Note: The coverage metrics \textbf{$TFC$} are computed within the \textbf{Cones of Influence (COI)} \cite{19} covered by the generated assertions, where a larger \textbf{COI} encompasses more signals. Therefore, in each design experiment, we use the number of signals involved in the largest \textbf{COI} as the standard denominator for the calculation.
\end{table}

In the experiment, we extract the correct assertions that pass FPV validation and use JasperGold once again to calculate the relevant coverage metrics. To more intuitively evaluate the capability of DeepAssert, we select four representative designs: I2C, ECG, Pairing, and SHA3. The detailed descriptions of these four designs are presented in Table \ref{Summary of Design}.

\begin{table}[h]
\centering
\renewcommand{\arraystretch}{1.4} 
\caption{Summary of Designs}
\label{Summary of Design}
\setlength{\tabcolsep}{3.5pt}
\begin{tabular}{c|c|c|c}
\hline
\hline
\cellcolor{gray!20}\textbf{Design Name} & \cellcolor{gray!20}\textbf{Func. Description} & \cellcolor{gray!20}\textbf{LoC} &
\cellcolor{gray!20}\textbf{Num. of Cells} \\ \hline
\makecell{\texttt{I\textsuperscript{2}C}} & \makecell{\hspace*{-4pt} Serial communication protocol.} & \makecell{5369} & \makecell{756}\\
\makecell{\texttt{SHA3}} & \makecell{\hspace*{-17pt} Hash function computation.} & \makecell{141185} & \makecell{22228}\\ 
\makecell{\texttt{ECG}} & \makecell{\hspace*{-12pt} Biological signal acquisition.} & \makecell{398686} & \makecell{59084}\\ 
\makecell{\texttt{Pairing}} & \makecell{\hspace*{-13pt} Cryptographic key exchange.} & \makecell{1561498} & \makecell{228287}\\ 
\hline
\hline
\end{tabular}
\end{table}

\subsection{Experimental Results}

\subsubsection{\textbf{Comparison of Deep Assertion Generation by DeepAssert and Previous Methods}}

Since both AssertLLM and Spec2Assertion rely on signal-related information from high-level specifications, they cannot directly generate deep assertions for modules. To ensure fair comparison, we replace signal names in their extraction process with module-relevant ones, forcing them to generate deep assertions. 

\begin{table}[h]
\centering
\caption{Performance of DeepAssert in generating deep assertions}
\label{Performance of DeepAssert}
\begin{tabular}{c|cccc}
\toprule
\rowcolor{gray!20} 
\textbf{Metric} & \textbf{I\textsuperscript{2}C} & \textbf{Pairing} & \textbf{ECG} & \textbf{SHA3} \\
\midrule
$N/S/P$ & 25/25/17 & 28/24/13 & 26/26/12 & 24/24/20 \\
$NVR$(\%) & \cellcolor{lightgreen}100 & \cellcolor{lightgreen}100 & \cellcolor{lightgreen}92.86 & \cellcolor{lightgreen}100 \\
$BFC$(\%) & \cellcolor{lightgreen}82.79 & \cellcolor{lightgreen}89.82 & \cellcolor{lightgreen}82.22 & \cellcolor{lightgreen}80 \\
$SFC$(\%) & \cellcolor{lightgreen}83.06 & \cellcolor{lightgreen}88.66 & \cellcolor{lightgreen}80 & \cellcolor{lightgreen}82.93 \\
$TFC$(\%) & 79.79 & 60.40 & 57.85 & 78.18 \\
\bottomrule
\end{tabular}
\end{table}

Results show that these methods struggle to produce effective deep assertions, as deep assertion generation is highly challenging. Prior methods near 0\% in related coverage metrics. For example, the forced deep assertions generated by AssertLLM for I2C are mostly ineffective, merely checking signal bit-width requirements without significantly improving coverage metrics, as shown below:
\begin{equation*}
\scriptsize
\begin{aligned}
&\texttt{assert property(\$bits(i2c\_master\_byte\_ctrl.ena)==1);}
\end{aligned}
\end{equation*} 
In contrast, DeepAssert surpasses 80\% coverage across most metrics with just a few deep assertions, as shown in Table \ref{Performance of DeepAssert}, so only coverage results of DeepAssert are displayed in the table.

\subsubsection{\textbf{Integration Experiments of DeepAssert with Previous Methods}}

\begin{table*}[h]
\centering
\caption{Performance comparison of AssertLLM and AssertLLM + DeepAssert}
\label{Performance comparison of AssertLLM and AssertLLM + DeepAssert}
\begin{tabularx}{\textwidth}{
  >{\centering\arraybackslash}X | 
  *{4}{>{\centering\arraybackslash}X|} 
  *{4}{>{\centering\arraybackslash}X} 
}
\toprule
\multirow{2}{*}{Metric} & \multicolumn{4}{c|}{\cellcolor{gray!20}\textbf{AssertLLM}} & \multicolumn{4}{c}{\cellcolor{lightblue}\textbf{AssertLLM + DeepAssert}} \\
\cline{2-9}
& I\textsuperscript{2}C & Pairing & ECG & SHA3 & I\textsuperscript{2}C & Pairing & ECG & SHA3 \\
\midrule
$N/S/P$ & 127/112/58 & 32/32/12 & 44/44/19 & 31/31/26 & 171/164/74 & 60/56/25 & 70/70/31 & 55/55/46 \\
$NVR$(\%) & 100 & 13.64 & 76.67 & 80 & 100 & \cellcolor{lightgreen}62 & \cellcolor{lightgreen}83.5 & \cellcolor{lightgreen}85 \\
$BFC$(\%) & 80.23 & 76.12 & 82.22 & 92 & \cellcolor{lightgreen}82.84 & \cellcolor{lightgreen}82.82 & 82.22 & \cellcolor{lightgreen}100 \\
$SFC$(\%) & 82.26 & 83.63 & 80.74 & 90.24 & \cellcolor{lightgreen}83.87 & \cellcolor{lightgreen}83.66 & 80.74 & \cellcolor{lightgreen}95.12 \\
$TFC$(\%) & 78.93 & 75.44 & 60.8 & 89.41 & \cellcolor{lightgreen}80.61 & 75.44 & \cellcolor{lightgreen}62.29 & 89.41 \\
\bottomrule
\end{tabularx}
\end{table*}

\begin{table*}[h]
\centering
\caption{Performance comparison of Spec2Assertion and Spec2Assertion + DeepAssert}
\label{Performance comparison of Spec2Assertion and Spec2Assertion + DeepAssert}
\begin{tabularx}{\textwidth}{
  >{\centering\arraybackslash}X | 
  *{4}{>{\centering\arraybackslash}X|} 
  *{4}{>{\centering\arraybackslash}X} 
}
\toprule
\multirow{2}{*}{Metric} & \multicolumn{4}{c|}{\cellcolor{gray!20}\textbf{Spec2Assertion}} & \multicolumn{4}{c}{\cellcolor{lightblue}\textbf{Spec2Assertion + DeepAssert}} \\
\cline{2-9}
& I\textsuperscript{2}C & Pairing & ECG & SHA3 & I\textsuperscript{2}C & Pairing & ECG & SHA3 \\
\midrule
$N/S/P$ & 90/89/49 & 45/41/15 & 35/29/19 & 42/42/28 & 115/114/66 & 73/65/28 & 61/55/31 & 66/66/48 \\
$NVR$(\%) & 100 & 100 & 100 & 92.24 & 100 & 100 & 100 & \cellcolor{lightgreen}97.83 \\
$BFC$(\%) & 87.87 & 83.58 & 79.51 & 90.89 & \cellcolor{lightgreen}89.51 & \cellcolor{lightgreen}83.82 & \cellcolor{lightgreen}83.12 & \cellcolor{lightgreen}96 \\
$SFC$(\%) & 89.44 & 66.67 & 81.05 & 87.78 & \cellcolor{lightgreen}91.05 & \cellcolor{lightgreen}80.66 & 81.05 & \cellcolor{lightgreen}92.68 \\
$TFC$(\%) & 87.85 & 51.04 & 78.55 & 78.31 & \cellcolor{lightgreen}88.55 & \cellcolor{lightgreen}67.19 & 78.55 & \cellcolor{lightgreen}78.36 \\
\bottomrule
\end{tabularx}
\end{table*}

To balance resource utilization and coverage, we generate a few deep assertions, adjustable via parameters. DeepAssert can integrate with existing methods post-Step 2, but we didn't pursue this to ensure deep assertion quality.

To explore whether DeepAssert can boost the coverage of prior methods, we integrated it with AssertLLM and Spec2Assertion. Table \ref{Performance comparison of AssertLLM and AssertLLM + DeepAssert} and Table \ref{Performance comparison of Spec2Assertion and Spec2Assertion + DeepAssert} shows the metric improvements before and after integration. The results indicate that even in large circuits, a few deep assertions from DeepAssert significantly enhance several coverage metrics, proving their high quality.

\subsubsection{\textbf{DeepAssert in Action via Mutation Testing for Error Coverage and Localization}} 

\begin{figure}[H]
\centering
\includegraphics[width=1.01\linewidth]{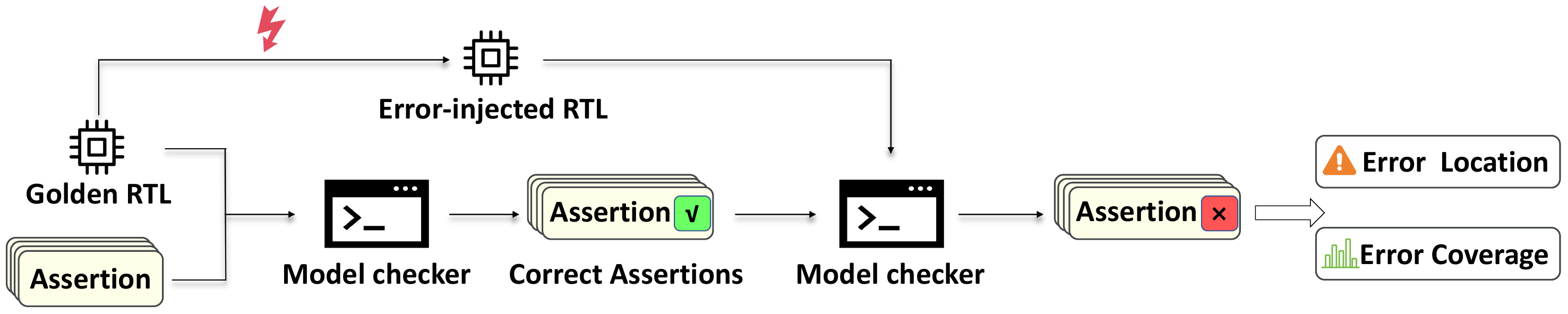}
\caption{Mutation testing process for assertion triggering.}
\label{fig:7}
\end{figure}

To further explore DeepAssert's performance in mutation testing, we utilize the JasperGold tool to inject mutation errors into four designs and performe error localization and coverage analysis, following the steps illustrated in Fig.\ref{fig:7}.

\begin{figure}[h]
\centering
\includegraphics[width=1.02\linewidth]{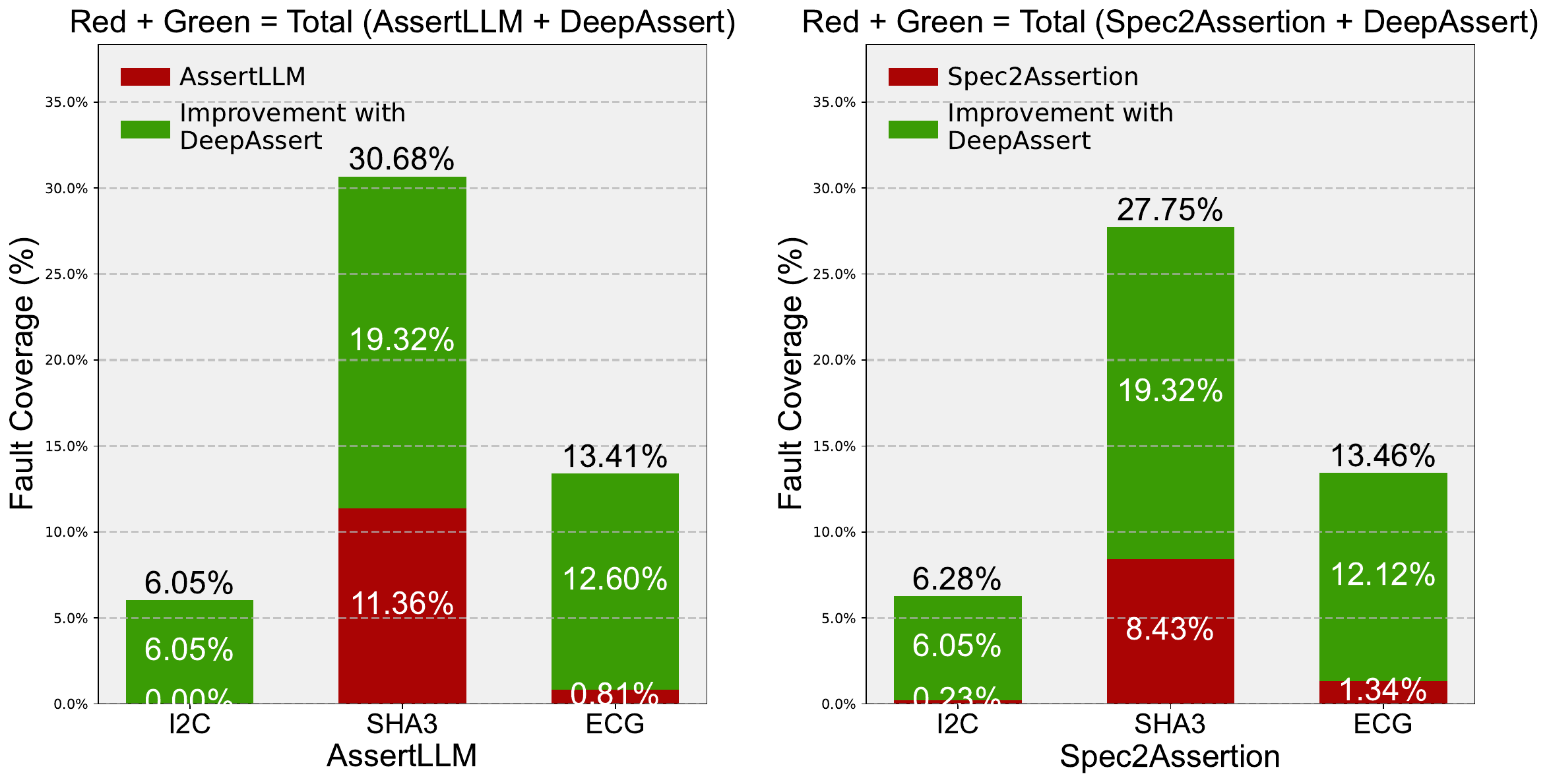}

\scriptsize 
\raggedright 
*Note: Due to the large scale of the Pairing, it would require over 120 hours for mutation testing, so we exclude the Pairing from the experiment.
\caption{Error Coverage Enhancement in Mutation Testing via Integration of AssertLLM and Spec2Assertion with DeepAssert.}
\label{fig:9}
\end{figure}

As Fig. \ref{fig:9} shows, integrating DeepAssert with the two prior methods greatly boosts error coverage in mutation testing, highlighting the effectiveness of deep assertions. In contrast, assertions from the other methods are often ineffective, yielding low error coverage or even 0\%.

\begin{figure}[h]
\centering
\includegraphics[width=0.85\linewidth]{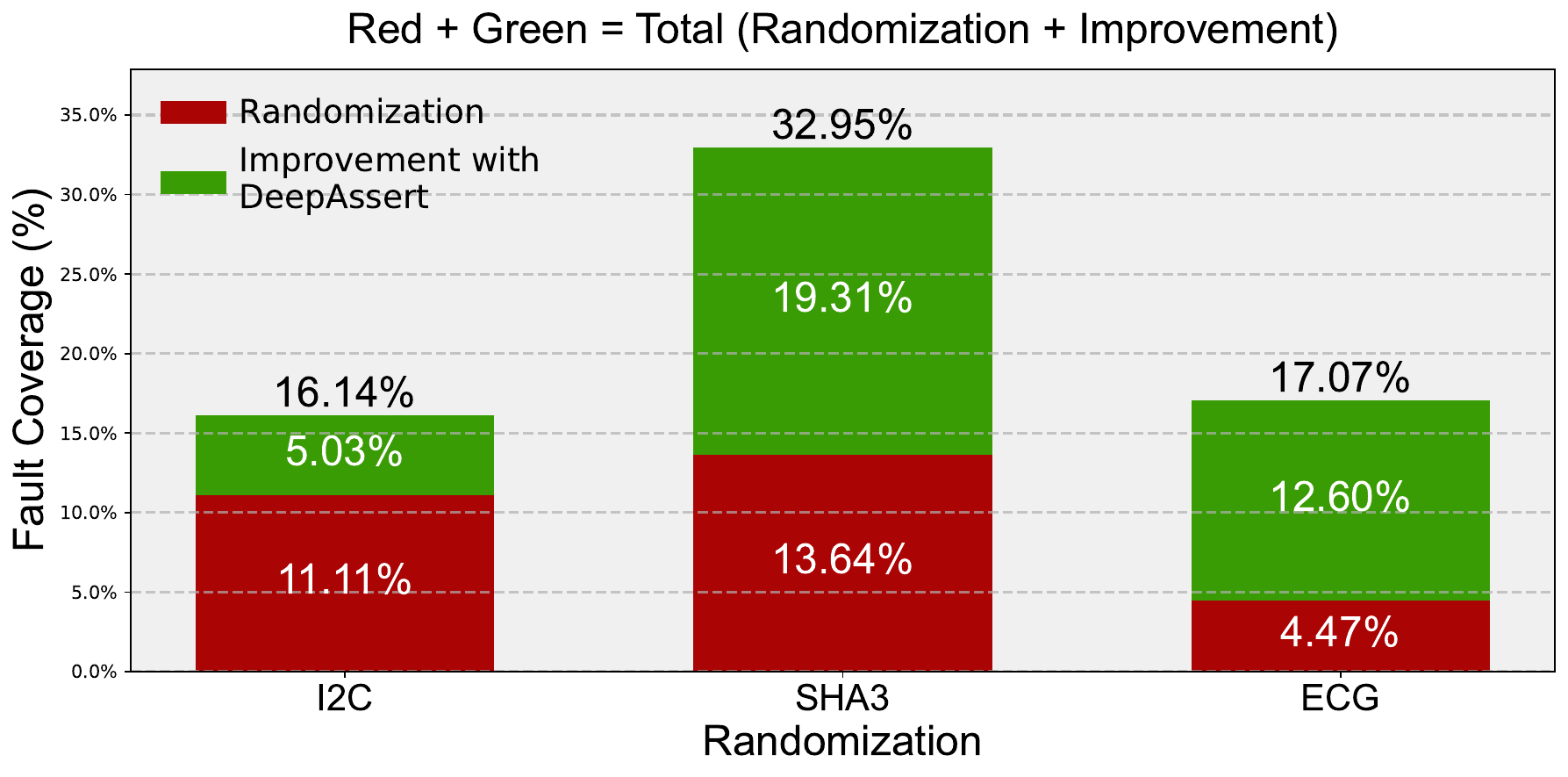}
\caption{Error Coverage Enhancement in Mutation Testing via Integration of Randomized Assertion with DeepAssert.}
\label{fig:12}
\end{figure}

To further assess the quality of Deep assertions, we randomly select key variables from four designs, craft high-quality manual assertions, and perform mutation testing. As Fig. \ref{fig:12} shows, DeepAssert's assertions still significantly improve mutation error coverage even when combined with high-quality manual assertions, further proving their effectiveness.

\begin{figure}[h]
\centering
\includegraphics[width=1\linewidth]{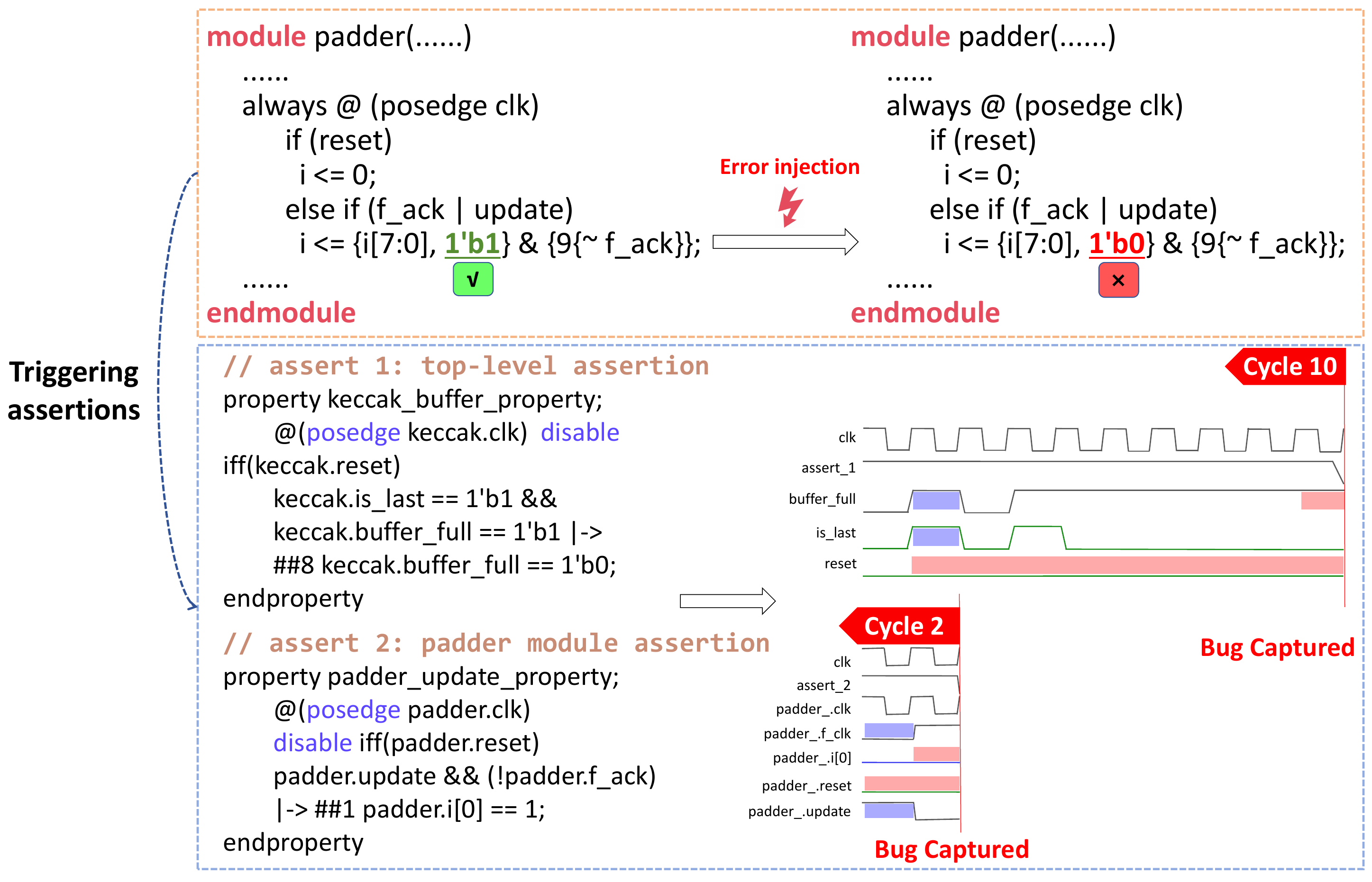}
\caption{A case of rapid assertion triggering and efficient debugging via assertion in the padder module of SHA3.}
\label{fig:10}
\end{figure}

Moreover, in our manual error injection attempts, we specifically activate an assertion in the top-level design and a deep assertion within a module, as shown in Fig. \ref{fig:10}. 

In this case study, injecting a single error into a module named \textit{padder} triggered both assertions simultaneously. The top-level assertion require ten clock cycles to trigger and could only infer an error's presence, while the module-level deep assertion trigger in just two clock cycles and directly identify the error signal, significantly reducing debugging time.

\section{Conclusion}

In conclusion, this paper introduces DeepAssert, an LLM-aided framework for generating deep assertions for modules in IC designs. By analyzing high-level specifications and module invocation relationships, DeepAssert extracts module-level specifications without relying on RTL implementation details, enabling fine-grained deep assertion generation. Experiments on OpenCores benchmarks show that DeepAssert is highly compatible, improving assertion quality and significantly reducing debugging time. Future work will focus on enhancing accuracy and robustness to handle more complex designs and further improve efficiency.

\bibliographystyle{plain} 
\bibliography{reference}

\end{document}